\documentstyle[epsfig]{mn}
\newif\ifAMStwofonts
\AMStwofontstrue

\def\cirx1{Cir$\,${X}-1}

\ifoldfss
  \ifCUPmtlplainloaded \else
    \NewTextAlphabet{textbfit} {cmbxti10} {}
    \NewTextAlphabet{textbfss} {cmssbx10} {}
    \NewMathAlphabet{mathbfit} {cmbxti10} {} 
    \NewMathAlphabet{mathbfss} {cmssbx10} {} 
  \fi
  \ifAMStwofonts
    \ifCUPmtlplainloaded \else
      \NewSymbolFont{upmath} {eurm10}
      \NewSymbolFont{AMSa} {msam10}
      \NewMathSymbol{\upi}     {0}{upmath}{19}
      \NewMathSymbol{\umu}     {0}{upmath}{16}
      \NewMathSymbol{\upartial}{0}{upmath}{40}
      \NewMathSymbol{\leqslant}{3}{AMSa}{36}
           
    \fi
  \fi
\fi 

\ifnfssone
  \newmathalphabet{\mathit}
  \addtoversion{normal}{\mathit}{cmr}{m}{it}
  \addtoversion{bold}{\mathit}{cmr}{bx}{it}
  \newmathalphabet{\mathbfit} 
  \addtoversion{normal}{\mathbfit}{cmr}{bx}{it}
  \addtoversion{bold}{\mathbfit}{cmr}{bx}{it}
  \newmathalphabet{\mathbfss} 
  \addtoversion{normal}{\mathbfss}{cmss}{bx}{n}
  \addtoversion{bold}{\mathbfss}{cmss}{bx}{n}
  \ifAMStwofonts
    \ifCUPmtlplainloaded \else
      \UseAMStwoboldmath
      \makeatletter
      \new@mathgroup\upmath@group
      \define@mathgroup\mv@normal\upmath@group{eur}{m}{n}
      \define@mathgroup\mv@bold\upmath@group{eur}{b}{n}
      \edef\UPM{\hexnumber\upmath@group}
      \new@mathgroup\amsa@group
      \define@mathgroup\mv@normal\amsa@group{msa}{m}{n}
      \define@mathgroup\mv@bold\amsa@group{msa}{m}{n}
      \edef\AMSa{\hexnumber\amsa@group}
      \makeatother
      \mathchardef\upi="0\UPM19
      \mathchardef\umu="0\UPM16
      \mathchardef\upartial="0\UPM40
      \mathchardef\leqslant="3\AMSa36
    \fi
  \fi
\fi 

\ifnfsstwo
  \DeclareMathAlphabet{\mathbfit}{OT1}{cmr}{bx}{it}
  \SetMathAlphabet\mathbfit{bold}{OT1}{cmr}{bx}{it}
  \DeclareMathAlphabet{\mathbfss}{OT1}{cmss}{bx}{n}
  \SetMathAlphabet\mathbfss{bold}{OT1}{cmss}{bx}{n}
  \ifAMStwofonts
    \ifCUPmtlplainloaded \else
      \DeclareSymbolFont{UPM}{U}{eur}{m}{n}
      \SetSymbolFont{UPM}{bold}{U}{eur}{b}{n}
      \DeclareSymbolFont{AMSa}{U}{msa}{m}{n}
      \DeclareMathSymbol{\upi}{0}{UPM}{"19}
      \DeclareMathSymbol{\umu}{0}{UPM}{"16}
      \DeclareMathSymbol{\upartial}{0}{UPM}{"40}
      \DeclareMathSymbol{\leqslant}{3}{AMSa}{"36}
    \fi
  \fi
\fi 

\ifCUPmtlplainloaded \else
  \ifAMStwofonts \else 
    \def\upi{\pi}
    \def\umu{\mu}
    \def\upartial{\partial}
  \fi
\fi

\title{Circinus X-1: survivor of a highly asymmetric supernova}
\author[T. M. Tauris, R. P. Fender, E. P. J. van den Heuvel,
        H. M. Johnston and K. Wu]
       {T. M. Tauris$^{1}$, 
        R. P. Fender$^{1}$, E. P. J. van den Heuvel$^{1}$, 
        H. M. Johnston$^{2}$ and K. Wu$^{3}$\\
       $^1$Center for High-Energy Astrophysics and
           Sterrenkundig Instituut ``Anton Pannekoek'', University of Amsterdam,
           Kruislaan 403, 1098-SJ Amsterdam, The Netherlands\\
       $^2$Anglo-Australian Observatory, P.O. Box 296, Epping,
           NSW 1710, Australia\\
       $^3$Research Centre for Theoretical Astrophysics, School of Physics,
           University of Sydney, NSW 2006, Australia}

\date{Accepted: 1999 September 2.  
      Received: 1999 March 18}

\pagerange{\pageref{firstpage}--\pageref{lastpage}}
\pubyear{1999}

\begin{document}

\maketitle

\label{firstpage}

\begin{abstract}
We have analyzed the kinematical parameters of \cirx1 to constrain
the nature of its companion star, the eccentricity of the binary
and the pre-supernova parameter space. We argue that the companion
is most likely to be a low-mass ($\la 2.0\,M_{\odot}$) unevolved star
and that the eccentricity of the orbit is $0.94\pm0.04$.
We have evaluated the dynamical effects of the supernova explosion
and we find it must have been asymmetric. 
On {\em average}, we find that a kick of $\sim$740~km~s$^{-1}$ is needed to account for
the recently measured radial velocity of +430~km~s$^{-1}$ 
(Johnston, Fender \& Wu) for this extreme system. 
The corresponding {\em minimum} kick velocity is $\sim$500~km~s$^{-1}$.
This is the largest kick needed to explain the motion of any
observed binary system. 
If \cirx1 is associated with the supernova remnant G321.9-0.3 then we find a
limiting minimum age of this remnant of $\sim\!60\,000$ yr.
Furthermore, we predict that the companion star has lost $\sim\!10\,$\% of its 
mass as a result of stripping and ablation from the impact of the supernova 
shell shortly after the explosion.
\end{abstract}


\section{Introduction}
\cirx1 is a unique binary in many aspects. Despite intensive {X}-ray
observations since its discovery in the early 1970s very little is
known about its stellar components besides an orbital period
of 16.6 days (Kaluzienski et al. 1976). It is still uncertain
whether the binary is a high-mass (HMXB) or low-mass {X}-ray binary (LMXB).
The discovery of Type~I {X}-ray bursts (Tennant et al. 1986) demonstrates
that the accreting compact object must be a neutron star.
In a recent paper Johnston, Fender \& Wu (1999) presented new optical
and infrared observations of \cirx1 which show asymmetric emission lines.
Combined with twenty years of archival {H}$\alpha$ line emission data,
they interpreted that the narrow components of the lines imply a radial
velocity of +430~km~s$^{-1}$ for the system (including a small correction
for the Galactic rotation). This is the highest velocity known for any
{X}-ray binary.\\
\cirx1 is located 25$^{\prime}$ from the centre of the supernova remnant
G321.9-0.3 and is apparently connected by a radio nebula (Haynes et al. 1986).
Furthermore, there is some evidence for an association from observations
of arcmin-scale collimated structures within the nebula (Stewart et al. 1993)
which are likely to originate from an arcsec-scale jet which has been 
observed (Fender et al. 1998) to be aligned with these larger structures.
If the suggested association is correct then \cirx1 is a young ($<10^5$ yr)
runaway system and, in combination with the recent estimated distance
to the remnant of 5.5~kpc (Case \& Bhattacharya 1998), the inferred minimum
transverse velocity (390~km~s$^{-1}$) yields a resulting 3{\small D} space
velocity of $>$580~km~s$^{-1}$. In this Letter we investigate what can be
learned about \cirx1 and the effects of the supernova explosion which
gave rise to this high runaway velocity and orbital period of 16.6 days.

\section{Dynamical effects of an asymmetric supernova explosion}
We use the analytical formulae by Tauris \& Takens (1998) to calculate
the dynamical effects of an asymmetric supernova (SN) in \cirx1. 
These formulae also include the effect of shell impact on the companion star.
We assume that the exploding {He}-star with mass $M_{\rm He}$
(the progenitor of the neutron star with mass $M_{\rm NS}$) is at the origin
of the cartesian coordinate system which we shall use in our description. 
The positive z-axis points in the direction of the pre-SN orbital angular
momentum. The positive y-axis points towards the companion star (with mass
$M_2$) and the positive x-axis points in the direction of
$\vec{v}$, which is the pre-SN relative velocity vector of the {He}-star
with respect to the companion star. 
We assume the pre-SN orbit is circular and $r_0$ is the pre-SN separation 
between the two stellar components.
The systemic velocity of a binary which survives the SN is:
\begin{equation}
  |\vec{v}_{\rm sys}|=
       \sqrt{\Delta p_{\rm x}^2+\Delta p_{\rm y}^2+\Delta p_{\rm z}^2}/
        (M_{\rm NS}+M_{\rm 2f})
\end{equation}
where the change in momentum resulting from the SN is:\\
\[
\Delta p_{\rm x}=\displaystyle\frac{M_{\rm NS}M_2-M_{\rm He}M_{\rm 2f}}{M_{\rm He}+M_2}\,v 
                  +M_{\rm NS}\,w\cos\vartheta \]
\vspace{-0.3cm}
\begin{equation}
\Delta p_{\rm y}=M_{\rm NS}\,w\sin\vartheta\cos\varphi 
                  +M_{\rm 2f}\,v_{\rm im} 
\end{equation}
\vspace{-0.3cm}
\[
\Delta p_{\rm z}=M_{\rm NS}\,w\sin\vartheta\sin\varphi\\
\]
A kick with magnitude $w$ is imparted to the neutron star (NS) by an asymmetric
explosion. The pre-SN relative orbital speed is $v=\sqrt{G(M_{\rm He}+M_2)/r_0}$
and the angle between $\vec{v}$ and $\vec{w}$ is $\vartheta$. The second
position angle $\varphi$ of $\vec{w}$, is taken such that
$w_y=w\sin\vartheta\cos\varphi$. 
Finally, $v_{\rm im}$ is an effective speed taking into account the combined
effects of incident shell momentum and the subsequent momentum
resulting from mass loss due to stripping and ablation of stellar
material from the surface layers heated by the passing shock wave.
It is given by (Wheeler, Lecar \& McKee 1975; see also Fryxell \& Arnett 1981):
\begin{equation}
  v_{\rm im} = \eta\,v_{\rm eject}\left(\frac{R_2 x_{\rm crit}}{2\,r_0}\right)^2
               \left(\frac{M_{\rm shell}}{M_2}\right)
               \frac{1+\ln(2\,v_{\rm eject}/v_{\rm esc})}
               {1-(F_{\rm strip}+F_{\rm ablate})}
\end{equation}
where $R_2$ is the initial radius of the companion star,
$v_{\rm eject}$ is the speed of the material ejected in the SN,
$v_{\rm esc}=\sqrt{2GM_2(x_{\rm crit})/R_2\,x_{\rm crit}}$ is the escape
velocity from the stripped companion star and
the parameter $x_{\rm crit}$ is a critical fraction of the radius outside
of which the total mass fraction, $F_{\rm strip}$ is stripped and inside
of which a certain fraction, $F_{\rm ablate}$ is ablated.
Thus after the shell impact the new mass of the companion is given by:
$M_{\rm 2f}=M_2 \, (1-(F_{\rm strip}+F_{\rm ablate}))$.
A correction to the assumption of plane parallel layers is expressed by $\eta$.
We assume $\eta =0.5$.\\
Wheeler, Lecar \& McKee (1975) have defined a parameter, $\Psi$
in order to evaluate $x_{\rm crit}$ and $F=F_{\rm strip}+F_{\rm ablate}$
(their tabulation is given in Table~1):
\begin{equation}
\Psi \equiv \left( \frac{R_2}{2r_0} \right)^2
            \left( \frac{M_{\rm shell}}{M_2} \right)
            \left( \frac{v_{\rm eject}}{v_{\rm esc}} -1 \right)
\end{equation}
In order to calculate the possible values of $\Psi$ for the case of \cirx1
we assume that a main-sequence companion star can be approximated by a
polytropic model with index $n$=3. In this case one has:
$M_2(x_{\rm crit})/R_2\,x_{\rm crit}\simeq 5/3\:M_{\odot}R_{\odot}^{-1}$
for $0.2 < x_{\rm crit} < 0.7$ (to an accuracy within 15\,\%).
Furthermore we assume 
$E_{\rm eject} = 1/2\;M_{\rm shell}\, v_{\rm eject}^2 = 1.0\times 10^{51}$ erg.\\
The absolute minimum ZAMS mass for producing a NS is $\sim\!8\,M_{\odot}$.
Such an early-type star evolves on a short timescale, and hence at the time
of the SN the maximum age of the companion is $\sim\!40$ Myr.
The companion star was less massive, and is therefore most likely to be unevolved.
We fitted a mass-radius relation for ZAMS stars using the
stellar evolution models of Pols et al. (1998):
\begin{equation}
  R_2/R_{\odot} = \left\{ \begin{array}{ll}
      0.86\,(M_2/M_{\odot})^{1.21}
               & \mbox{\hspace{0.7cm} $M_2 < 1.5\,M_{\odot}$}\\
      1.15\,(M_2/M_{\odot})^{0.50}
               & \mbox{\hspace{0.7cm} $M_2 > 1.5\,M_{\odot}$}
      \end{array}
         \right.
\end{equation}
assuming {X}=0.70, {Z}=0.02 and a convective mixing-length parameter $\alpha=2.0$.\\
   \begin{table}
     \caption{Parameters as a function of $\Psi$ (see text) for a
              companion with the structure of a polytrope with index $n$=3.}
     \begin{flushleft}
        \begin{tabular}{lllll}
           \hline\noalign{\smallskip}
           $\Psi$ & $x_{\rm crit}$ & $F_{\rm strip}$ & $F_{\rm ablate}$ & $F$\\
           \noalign{\smallskip}
           \hline\noalign{\smallskip}
          10 & 0.17 & 0.73 & 0.29 & $\sim1$\\
           5 & 0.26 & 0.43 & 0.34 & 0.77\\
           4 & 0.29 & 0.35 & 0.34 & 0.69\\
           2 & 0.37 & 0.19 & 0.28 & 0.47\\
           1 & 0.44 & 0.10 & 0.19 & 0.29\\
         0.4 & 0.52 & 0.04 & 0.11 & 0.15\\
         0.2 & 0.58 & 0.02 & 0.07 & 0.09\\
         0.1 & 0.64 & 0.01 & 0.04 & 0.05\\
        0.01 & 0.78 & 0.001 & 0.006 & 0.01\\
       0.001 & 0.87 & $\sim0$ & $\sim0$ & $\sim0$\\
          \noalign{\smallskip}
          \hline
        \end{tabular}
     \end{flushleft}
     \begin{list}{}{}
       \item[] This table was taken from Wheeler, Lecar \& McKee (1975).
     \end{list}
   \end{table}
The post-SN separation and eccentricity are given by:
\begin{equation}
  a=\frac{r_0}{2-\xi} \qquad \qquad e=\sqrt{1+(\xi-2)\,(Q+1)}
\end{equation}
where (see Tauris \& Takens 1998):
\begin{equation}
  \xi \equiv \frac{v^2+w^2+v_{\rm im}^{2}+
        2w\,(v\cos\vartheta-v_{\rm im}\sin\vartheta\cos\varphi)}{\tilde{m}v^2}
\end{equation}
\begin{equation}
  Q \equiv \xi -1 -
           \frac{(w\sin\vartheta\cos\varphi-v_{\rm im})^2}{\tilde {m} v^2}
\end{equation}
and $\tilde{m}\equiv (M_{\rm NS}+M_{\rm 2f})/(M_{\rm He}+M_2)$.\\
Prior to the SN we require that the companion star is able to fit 
{\em inside} its Roche-lobe, and after the SN it is required
that the companion star {\em must} fill its Roche-lobe
near the post-SN periastron, $a\,(1-e)$ in order to explain the
observed {X}-ray emission (which is caused by mass accreted onto
the neutron star from a disk which has to be fed by stellar material).
Combining our mass-radius relation with Eggleton's (1983) expression
for the Roche-lobe radius, $R_{\rm L}$ yields the following
criterion for the radius of the companion star:
\begin{equation}
  \left. R_{\rm L}(q_0)\right|_{r=r_0} \geq R_2 \geq R_{\rm 2f} > 
  \left. R_{\rm L}(q_{\rm f})\right|_{r=a(1-e)}
\end{equation}
where $R_{\rm L}(q)=r\,0.49\,q^{2/3}\,(0.6\,q^{2/3}+\ln(1+q^{1/3}))^{-1}$ and
the mass ratios before and after the SN are given by:
$q_0\equiv M_{\rm 2}/M_{\rm He}$ and $q_{\rm f}\equiv M_{\rm 2f}/M_{\rm NS}$,
respectively.

\section{Simulation results and discussion}  
\begin{figure*}
 \begin{center}
   \epsfig{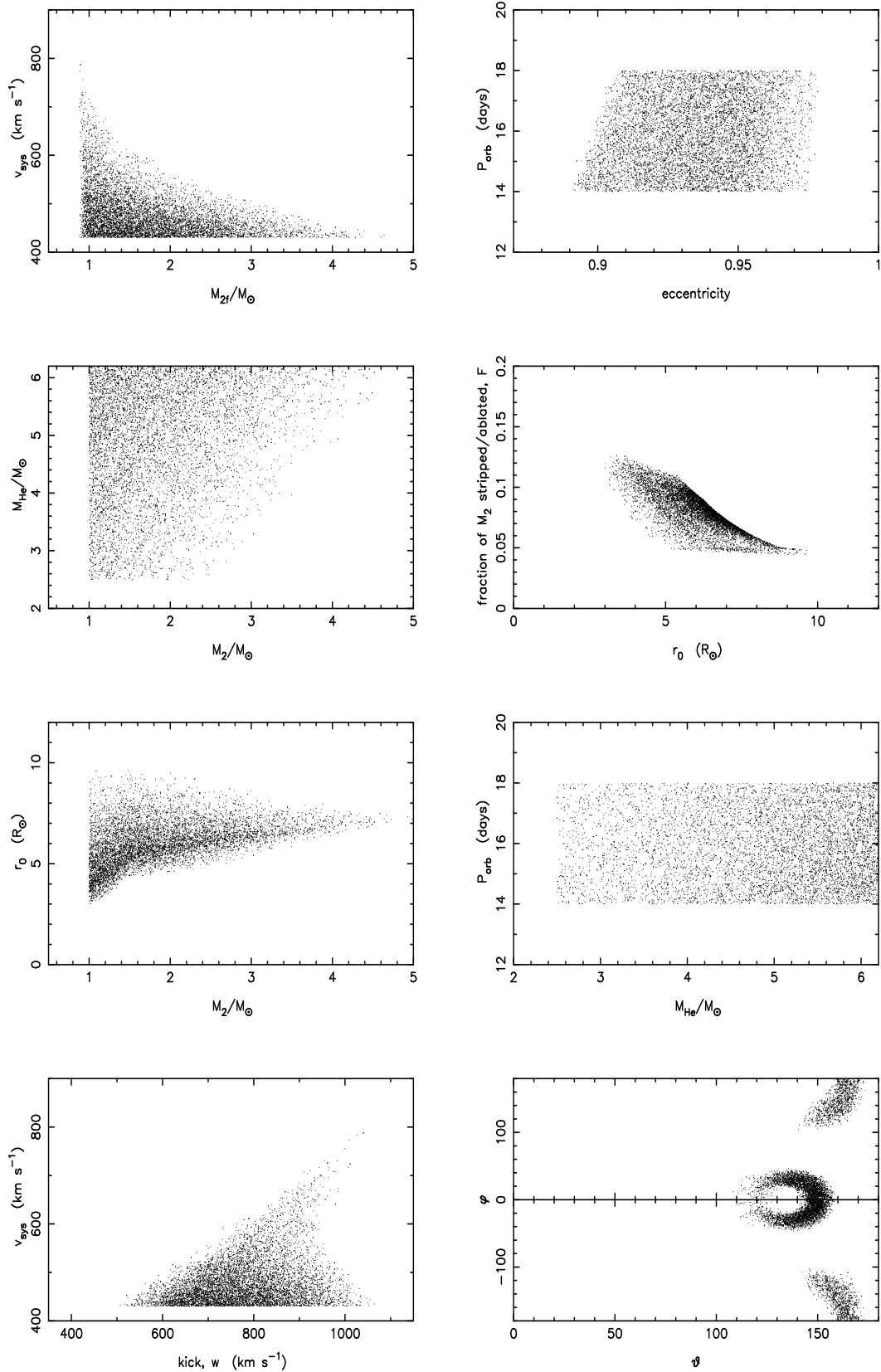}
 \end{center}
 \caption{The distribution of various pre- and post-SN binary parameters
obtained by Monte Carlo simulations in order to construct systems
similar to \cirx1. We assumed $M_{\rm NS}=1.3\,M_{\odot}$, $M_2 > 1.0\,M_{\odot}$ and 
$2.5 < M_{\rm He}/M_{\odot} < 6.5$. 
We chose flat trial distributions of $M_{\rm He}, M_2, r_0$ and $w$, and
assumed an isotropic distribution of kick directions.
The figure shows the characteristics of $10^4$ binaries
(out of $\sim\!3.4\times10^8$ trial systems) which passed our selection
criteria for resembling \cirx1 -- see text for discussions.
\label{fig1}}
\end{figure*}
   \begin{table*}
     \caption{The mean and minimum kick velocity imparted to the 
              neutron star as a function of the assumed systemic velocity of \cirx1.
              We assume $d$=5.5 kpc and $\delta\theta$=25$^{\prime}$ for G321.9-0.3.
              $v_{\rm sys}^2=v_{\rm rad}^2+v_{\perp}^2$ and
              $v_{\perp}=d\,\delta\theta/t_{\rm SNR}$. 
              All velocities are in km~s$^{-1}$ -- see also Fig.~2.}
     \begin{flushleft}
        \begin{tabular}{rrrl}
           \hline\noalign{\smallskip}
           $v_{\rm sys}$ & $\langle w \rangle$ & $w_{\rm min}$ & \\
           \noalign{\smallskip}
           \hline\noalign{\smallskip}
 430 & 740 & 500 & no association with G321.9-0.3 and $v_{\perp}=0$ is assumed (devils advocate's solution for lowest possible kick)\\
 430 & 740 & 520 & no association with G321.9-0.3, no shell impact effects included and $v_{\perp}=0$\\ 
 580 & 840 & 760 & association with G321.9-0.3 and $t_{\rm SNR}$=10$^5$ yr\\
 580 & 850 & 680 & association with G321.9-0.3, $t_{\rm SNR}$=10$^5$ yr and $E_{\rm eject} = 5.0\times 10^{51}$ erg.\\
 700 & 940 & 920 & association with G321.9-0.3 and $t_{\rm SNR}$=70\,000 yr\\
          \noalign{\smallskip}
          \hline
        \end{tabular}
     \end{flushleft}
   \end{table*}
We have performed Monte Carlo simulations of the SN effects on a large number of
binaries to determine the mass of the companion star, the eccentricity 
of the present orbit and the pre-SN orbital parameter space.
The constraints which must be fulfilled by any post-SN binary in order
to be consistent with present observations of \cirx1, are:
{I}) the companion star must fill its Roche-lobe at periastron,
{II}) the system must have a high
velocity of at least 430~km~s$^{-1}$ (Johnston, Fender \& Wu 1999)
and {III}) the system must have an orbital period between 14 and 18 days.
For \cirx1\\
$P_{\rm orb}=16.6$ days, but tidal effects (increasing
$P_{\rm orb}$) and accretion onto the NS (decreasing $P_{\rm orb}$ if
$M_{\rm 2f} > M_{\rm NS}$) could have affected the orbital evolution
since the SN\footnote{We tried to determine $\dot{P}_{\rm orb}$
from the All Sky Monitor XTE data of \cirx1 over the last 3 years
without any success. This is attributed to the poor timing resolution of
the data near periastron. (However Fender 1997, found some
evidence that the quadratic ephemeris is better than the linear,
implying $\dot{P}_{\rm orb}\neq 0$).}.
However, we are not able to trace back the detailed evolution since the SN.\\
The results of our simulations are shown in Fig.~1. 
We now summarize what can be learned from these calculations.

The top panel shows $v_{\rm sys}$ as a function of $M_{\rm 2f}$,
and $P_{\rm orb}$ as a function of eccentricity.
There is 76$\,$\% probability that $430 < v_{\rm sys}< 500$~km~s$^{-1}$
and only 4.5$\,$\% probability that $v_{\rm sys}> 580$~km~s$^{-1}$ if 
nothing is assumed to be known about the transverse velocity.
The most probable present mass of the companion star, $M_{\rm 2f}$ strongly 
decreases with increasing values of $M_{\rm 2f}$.
If $M_2 > 1.0\,M_{\odot}$ we find there is only 30$\,$\% and 5.6$\,$\%
probability that $M_{\rm 2f}> 2.0\,M_{\odot}$ and $3.0\,M_{\odot}$, respectively. 
We therefore conclude that \cirx1 is most likely to be an LMXB.
An absolute upper limit for $M_{\rm 2f}$ is $4.6\,M_{\odot}$.\\
In the right-hand panel we see that we can constrain $e=0.94\pm 0.04$.
The reason of this high expected eccentricity is that (as we have argued)
the companion must be unevolved and therefore has to pass by 
very close to the NS in order to transfer mass at each passage. 
(Also the extreme velocity of \cirx1 infers that the SN nearly
disrupted the system and thus $e$ is expected to be large).

In the second panel we plotted in the left-hand side the
distribution of the allowed masses of the two stellar
components prior to the SN. This, somewhat surprising plot, shows
that we can not constrain the combination of these two pre-SN masses very well.
All the combinations of masses shown can result in a post-SN
binary which resembles \cirx1. This result reflects the enormous
amount of freedom due to the unknown direction and magnitude
of the kick imparted to the newborn neutron star.
It should be noted that while the probability distribution of $M_2$ 
is decreasing with $M_2$,
the most likely mass of the exploding {He}-star
increases monotonically with $M_{\rm He}$.
It is seen that the distribution of $M_{\rm 2f}$ is slightly shifted towards
lower masses compared to the distribution of $M_2$. This reflects the mass loss 
from the companion star due to the shell impact.
This effect is shown in the right-hand side of the second panel, where we have plotted
the fraction of material lost from the companion star immediately after the SN 
as a function of the pre-SN separation, $r_0$. We see that we expect
5--12$\,$\% of the mass of the companion star to have been lost 
as a result of the shell impact. This is significant and future observations
of the supernova remnant might be able to give evidence for this
scenario. It is seen that the effect of the shell impact decreases
with $r_0$ as expected, see e.g. equation~(4).

In the third panel we plotted $r_0$ as a function of $M_2$, and
$P_{\rm orb}$ as a function of $M_{\rm He}$. The lower limit to $r_0$ as
a function of $M_2$ is given by the requirement that the companion
star must be able to fit inside its pre-SN Roche-lobe, cf. equation~(9). It can be seen
that it is difficult to produce a system like \cirx1 if $r_0$ is larger
than 9$\,R_{\odot}$ (if $r_0$ is larger it becomes very difficult to
keep the system bound after the SN and still require a large space velocity).
Such a tight pre-SN binary is most likely to originate from a common envelope
evolution. This fact also yields some preference for a low value of $M_2$, since
common envelopes are most likely to form for extreme mass ratios between
its stellar constituents (e.g. Iben \& Livio 1993). 

In the bottom panel we plotted $v_{\rm sys}$ as a function of the kick
velocity (actually {\em speed}) and the distribution of allowed kick angles,
in the left- and right-hand side, respectively.
Cumulative probability curves for the kick velocity, $w$ are presented in Fig.~2.
An important result is that under {\em no} circumstances, for any pre-SN
parameters, could \cirx1 be formed in a symmetric SN. In recent papers
Iben \& Tutukov (1996,1998) claimed that there is no proof for asymmetric SN 
in nature. Here our simulations show that the remarkably high radial velocity
observed in \cirx1 can only be explained by a substantial asymmetry in the SN
explosion.
For an average set of pre-SN parameters presented in Fig.~1
the probability for a binary to survive such a large kick is
less than 10$\,$\%.
The allowed parameter space of the kick angles show that the kick
must have been directed backwards ($\vartheta > 110\degr$)
with respect to the pre-SN
orbital motion of the exploding {He}-star. This makes sense since a backward
kick strongly increases the probability of a binary to stay bound 
(Flannery \& van den Heuvel 1975; Hills 1983). 
Similarly, we also expect the NS to be kicked in a direction close to the 
orbital plane ($\varphi\simeq 0\degr$ or $\varphi\simeq 180\degr$)
in order for the system to minimize its orbital energy and avoid disruption.
However, if the resulting velocity
vector of the NS relative to the companion star, just after
the shell decouples gravitationally from the system 
($\vec{u}_0=\vec{v}+\vec{w}-\vec{v}_{\rm im}$; Tauris \& Takens 1998)
is directed exactly towards the companion, then the NS is shot directly into 
the envelope of the companion star and the binary is assumed to
merge since $a(1-e)<R_{\rm 2f}$.
This explains the nice `shadow image' of the companion star
seen in the plot. Hence, in the case of \cirx1
the NS must have been kicked in a direction close to the companion star,
but without hitting it, in order for the binary to survive.
It is interesting to notice that the radio image of \cirx1 
(Stewart et al. 1993) seems to indicate that the jet is perpendicular
to the motion of the system. Since this jet is likely to be perpendicular
to the inner accretion disk and the orbital plane, the motion of the
binary seems to be in a direction along its orbital plane.
This supports the requirement that the kick was directed near
the pre-SN orbital plane.

In Table~2 we list the mean and minimum values of the kick velocity
needed to produce \cirx1 as a function of $v_{\rm sys}$.
It is interesting to notice 
that the minimum kick velocity needed to explain
the motion of \cirx1 is smaller when the effects of the shell impact
are included. This is explained by the fact that the momentum from the
incident shell ejecta blows the companion star in the same direction
as the NS is kicked, since the majority of NS are kicked in the direction of the
companion in order to survive (as mentioned above). Hence a smaller kick is needed to 
produce a high runaway velocity, and the probability of surviving
the SN is also increased. 
\begin{figure}
 \begin{center}
   \epsfig{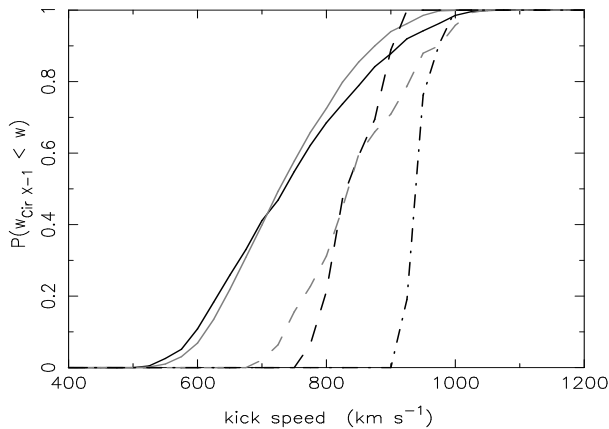}
 \end{center}
 \caption{Curves representing the probability that the kick imparted
to the NS in \cirx1 was less than the value given on the x-axis.
The full, dashed and dashed-dotted line correspond to
$v_{\rm sys}=430, 580$ and 700~km~s$^{-1}$, respectively.
The full gray line assumes no shell impact on the companion star and 
$v_{\rm sys}=430$~km~s$^{-1}$; the dotted gray line includes shell impact
with $E_{\rm eject}=5.0\times 10^{51}$ erg and assumes 
$v_{\rm sys}=580$~km~s$^{-1}$.
\label{fig2}}
\end{figure}

\subsection{Observations of the periastron passage}
We have shown for the condition of Roche-lobe overflow at periastron,
that \cirx1 is expected to have a highly eccentric orbit
($e=0.94\pm 0.04$). Therefore the companion star passes by the NS at
periastron in a very short interval of time. We find the fraction of the
orbital period the companion spends between orbital phases $\theta _1$
and $\theta _2$ is given by:
\begin{equation}
  \frac{\delta P_{\rm orb}}{P_{\rm orb}} = 
     1 - \frac{1}{2\pi}\,\left[ \frac{\sqrt{1-e^2}\,e\sin\theta}{1-e\cos\theta}
      +\tan ^{-1}\sqrt{\frac{1+e}{1-e}}\,\tan\frac{\theta}{2}
                        \right]^{\theta _2}_{\theta _1}
\end{equation}
\cirx1 has an orbital period of 16.6 days. Assuming $e$=0.94 we
find it only takes $\sim$3.5 hours for the companion to move
from $\theta_1 =270\degr$ to $\theta_2 =90\degr$ -- see Fig.~3.
The companion spends more than 99$\,$\% of its time between
$90\degr < \theta < 270\degr$. Hence observations must be planned
very carefully in order to catch the periastron passage
where e.g. the effects of companion heating due to {X}-ray irradiation
are most severe.
\begin{figure}
 \begin{center}
   \epsfig{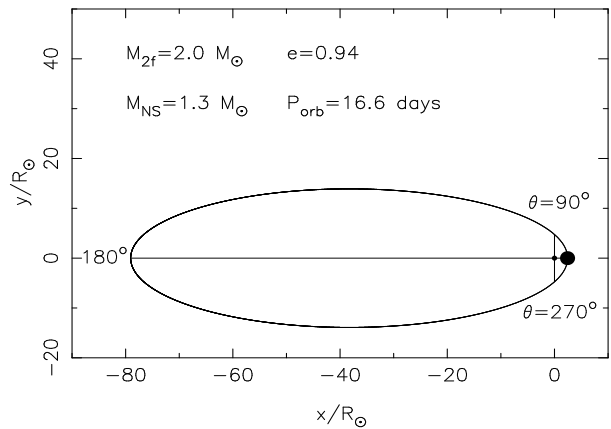}
 \end{center}
 \caption{Orbital geometry of \cirx1 in a reference frame
          fixed on the neutron star. We assumed 
          $M_{\rm 2f}=2.0\,M_{\odot}$ and $e=0.94$.
          The periastron passage $270\degr < \theta < 90\degr$
          only lasts $\sim$3.5 hours.
\label{fig3}}
\end{figure}

\subsection{Possible detection of ablated material from the 
shell impact and dependence on $E_{\rm eject}$}
Our calculations show that we expect $\sim$10$\,$\% of the mass
of the companion star to have been evaporated. Since $1.0 < M_2/M_{\odot} < 5.0$
we estimate that $\sim 0.1-0.5\,M_{\odot}$ of material in the SN nebula
originates from the {H}-rich envelope of the companion star. 
This quantity is equivalent to $\sim$10--25$\,$\% of
the amount of {He}-rich material which was
ejected from the collapsing {He}-star progenitor of the NS.
We therefore encourage detailed observations of the supernova
remnant G321.9--0.3 to search for evidence of the shell impact.
If $E_{\rm eject}$ is assumed to be $5.0\times 10^{51}$ erg then the
mean fraction of evaporated material increases to 15$\,$\%.
The constraints on the other parameters derived here are only slightly
affected by a similar increase in $E_{\rm eject}$ --  e.g. the
minimum required kick velocity is then 470~km~s$^{-1}$ instead of
500~km~s$^{-1}$. 

\subsection{On the age of the SN remnant G321.9-0.3}
Our simulations show that the maximum runaway velocity of \cirx1
is: $v_{\rm sys}^{\rm max}=800$ km s$^{-1}$. Hence the maximum
transverse velocity is constrained to be:
$v_{\perp}^{\rm max}=675$ km s$^{-1}$ (given $v_{\rm rad}=430$ km s$^{-1}$).
Since the transverse velocity is inversely proportional to the
age, $t_{\rm SNR}$ of G321.9-0.3 we therefore derive a minimum age
of this remnant of $t_{\rm SNR}^{\rm min}\simeq 60\,000$ yr,
if \cirx1 is associated with it. 

\section*{Acknowledgments}
T.M.T. and R.P.F. acknowledge the receipt of a Marie Curie Research Grant
from the European Commission. K.W. acknowledges the support from the
Australian Research Council through a fellowship and an ARC grant. 
We thank Jeroen Homan for looking into the archival XTE data.

{}
\end{document}